\documentclass[12pt]{article}
\usepackage[utf8]{inputenc}
\usepackage[english]{babel}
\usepackage[utf8]{inputenc}
\usepackage{amsmath}
\usepackage{float}
\usepackage{tcolorbox}
\usepackage{graphicx}
\usepackage[margin=.5in]{geometry}
\usepackage[colorinlistoftodos]{todonotes}
\usepackage{amsmath,amssymb,dsfont,amsthm}
\usepackage{graphicx}
\usepackage{dsfont}
\usepackage{listings}	
\usepackage{color}
\usepackage{bm}
\usepackage{wasysym}
\usepackage[colorinlistoftodos]{todonotes}
\usepackage{mathtools}
\usepackage{caption}
\usepackage{subcaption}
\usepackage{hyperref}
\usepackage{gensymb}
\title{To be or not to be? A spatial predictive crime model for Rochester}
\author{Ankani Chattoraj*, Rupam Acharyya*, Sabyasachi Shivkumar*, \\Md Iftekar Tanveer*,\\ Mohammad Rafayet Ali*}

\date{\today}
\begin{document}
\maketitle
\begin{abstract}
This project uses a spatial model (Geographically Weighted Regression) to relate various physical and social features to crime rates. Besides making interesting predictions from basic data statistics, the trained model can be used to predict on the test data. The high accuracy of this prediction on test data then allows us to  make predictions of crime probabilities in different areas based on the location, the population, the property rate, the time of the day/year and so on. This then further gives us the idea that an application can be built to help people traveling around Rochester be aware when and if they enter crime prone area. 
\end{abstract}

\section{Introduction}
Crime has been previously shown to have both temporal and spatial characteristics \cite{FELSON2003595,MONTOLIO201699}. Understanding such patterns can help fight crime better and ensure increased safety of the civilians of the region. 
Exploring the spatial patterns of crime can give us insight about the causal factors that drive them. Also different predictors may act differently in different regions of the city; meaning the same factor may have different weightage in explaining a certain type of crime in two different areas. A model which respects these spatial relations would allow us to better model the crime distribution and play a key role in understanding how crime can be controlled and prevented. 


The purpose of this study is to understand the crime patterns in the city of Rochester. Crime has previously been associated with different factors like weather~\cite{cohn1990weather}, stock value~\cite{fiorentini1997economics}, income~\cite{deutsch1992crime} in the area~\cite{deutsch1992crime}, population diversity etc. Taking all these factors into consideration our first objective was to selectively highlight interesting observations directly from the data statistics. This helped us identify what time of the day is more sensitive to crime; which month of the year has higher crime rate and also which areas in Rochester are safer than others.

In the second part of our data exploration, we built a model that can be trained on various physical and social features to predict their weighted contribution into crime of a region. However we did not want to use global regression models since variations over space that might exist in the data are suppressed in such models. So instead we used Geographically Weighted Regression model~\cite{brunsdon1996geographically}, that has been used to model crime data before~\cite{cahill2007using,wheeler2009comparing,arnio2012demography}. GWR uses local equations for every location in the dataset and gives us the weights for each feature affecting crime in that area. The model after being trained; works extremely well on the test set as can be seen from figures.

In the third part of the project deals with building a website such that based on these spatial predicts we can spread awareness. A person can be notified based on the time of the year/day, the weather and the place to which they are traveling that whether they are entering a crime prone area or not. They can be warned accordingly and made aware.

\section{The Data}\label{data_section}
For the modeling we have chosen the crime data of Rochester Police Department. This can be  \href{http://data-rpdny.opendata.arcgis.com/}{found here}. In this data for each crime the relevant informations such as the location of crime, time of the crime, type of crime etc. are also there. One of the information in this data is the crime type. But for our modeling purpose we have created an ordinal variable for each crime type. For example, we use the dependent  variable\textit{ \textbf{IsLarceny}} to denote the probability of larceny happening in a  GeoID (which is a unique identification number of a small geographical area).  We have introduced similar variables for each of the crime types. The goal of the model is to predict this probabilities for each GeoID. To add more features (or alternatively independent variable) along with this we have concatenated the 
demographic data of Rochester which can be found in \href{www.cityofrochester.gov}{www.cityofrochester.gov}. We extract the spatial features from this data such as: population density, property rates, ethnicity distribution for each GeoID (which is a unique identification number of a small geographical area).  On top of these we added a weather data found in \href{http://w2.weather.gov/climate/}{http://w2.weather.gov/climate/}. From this data we used the features like average temperatures, snowfall etc. This features act as temporal features of the model. 

\section{Modeling Approach}
\subsection{Observations}
Basic statistical analysis on the data led to some interesting observations.\\
We find that the time midnight is one of the most crime sensitive times of the day; most likely criminals assume that residents are asleep during that time and there will be less pedestrians on roads hence easy to rob lonely passerby or break into someone's house without being noticed. The peak at noon can be justified by the fact that most residents are at work and hence burglary, motor vehicle theft and larceny has a peak during those times.\\ We also find that crime rate increases during the beginning of winter and summer. This could be because during the winter it might get hard to commit crimes, and during summer, residents are away on vacations or leave the house more often and hence burglary and theft increases. There are more tourists also around the summer and fall and hence theft and larceny increases. \\For regions where the property price is high, occurrences of burglary is less but larceny is more. This can be explained with the idea that residents who can afford high property prices will have expensive houses equipped with advanced alarm systems and hence it will be hard to break into such houses, hence there is less burglary. Also, in high property price areas there are fewer individuals with low or no income, which explains why the crime rate is comparatively low.
\subsubsection{Midnight and Noon are high crime occurrence time}
In the past time of the day was found to be a good indicative variable of crime \cite{FELSON2003595}. \cite{MONTOLIO201699} showed that crime increases in certain times of the day. This is why we looked at the crimes happened at different time of the day. Figure 1 shows different type of crime happened from 2011 to 2017. The x-axis represents the time of the day in hours and the y-axis represents the percentage of crime. It is evident that during 12pm (Noon, around when people are away from work) the crime rate is at its peak. Also, there is a high percentage of crime happened at midnight (00:00hr); likely when residents sleep. The overall crime rate dropped at 5am (see Figure ).  It should be noted that the police shift change times are 7am (07:00), 3pm (15:00), and 11pm (23:00). It is interesting that the crime has almost no correlation with the police time shift changes (Pearson $r < 0.2$). Our finding aligns with the past findings that the probability of crime  increases around midnight \cite{MONTOLIO201699}. However, murder (see )does not show any unique patterns. This also aligns with the past findings from \cite{PEREIRA2016116}. 
\begin{figure}[H]
	\centering
    \includegraphics[scale=0.4]{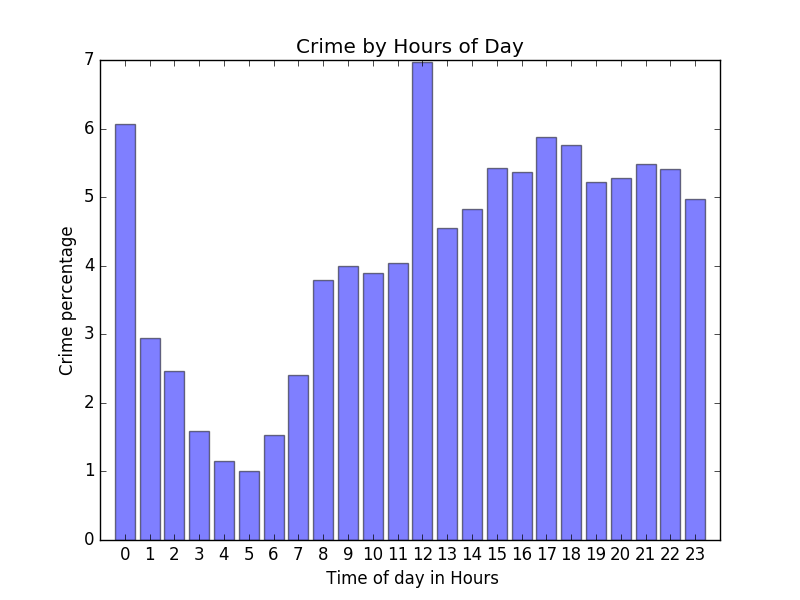}
    \label{fig:crimetime}
	\caption{Crime by time of day}
\end{figure}
\subsubsection{Temperature is correlated with crime rate}
\begin{figure}[H]
	\centering
    \includegraphics[scale=0.4]{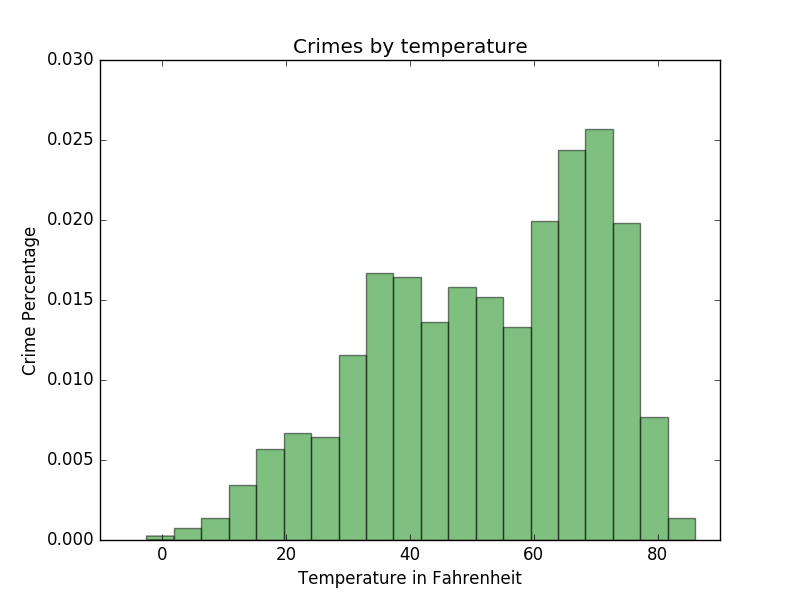}
	\caption{Crime by temperature}
    \label{fig:ALLCrime_by_temperature}
\end{figure}

Effects of temperature and climate change on crime has been studied in past \cite{Cotton86,mares2013climate,field1992effect}. We also looked at different crimes by temperature. Figure \ref{fig:ALLCrime_by_temperature} shows the histogram of the crime by temperature. We found that there are two peak points in the histogram. One was found around $32^{\circ}$ Fahrenheit and the other one is around $70^{\circ}$ Fahrenheit. The peak at $32^{\circ}$ Fahrenheit represents the crimes occurred during the beginning and the end of winter period. This can be motivated by the fact that its the holiday season. We will further investigate what event correlates with the crime more in later sections of this report. We also see a peak at $70^{\circ}$ Fahrenheit. This means that generally the crime is more likely to happen during the summer season. 
\subsubsection{Crime at its peak during Summer and Fall}
\begin{figure}[H]
	\centering
    \begin{subfigure}[b]{0.5\textwidth}
        \includegraphics[scale=0.5]{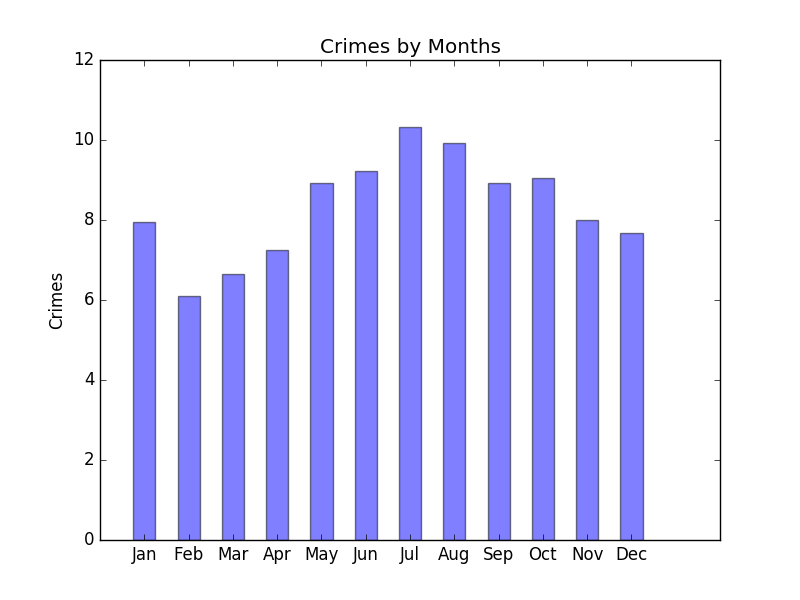}
        \caption{Crime by Months}
        \label{fig:ALLCrime_by_Months}
    \end{subfigure}%
    \begin{subfigure}[b]{0.5\textwidth}
        \includegraphics[scale=0.5]{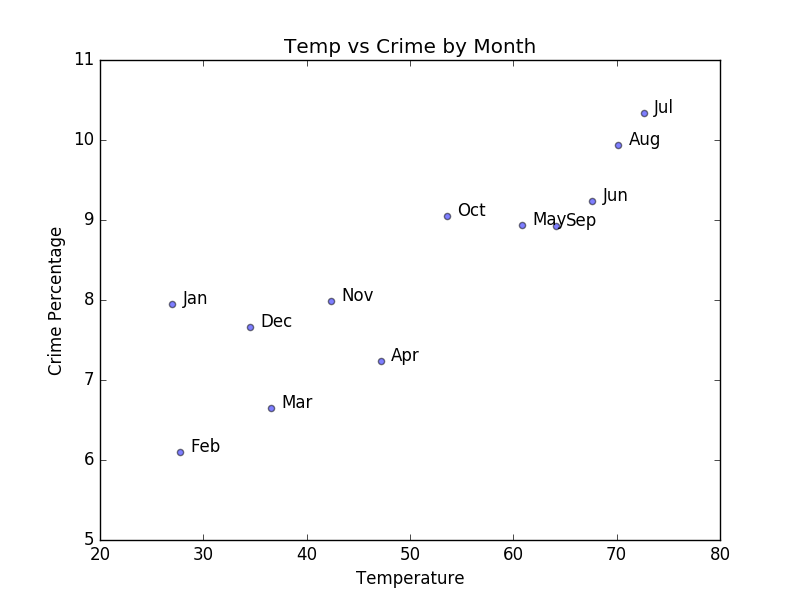}
		\caption{Crime vs Temperature by Months}
        \label{fig:Temp_vs_Crime_by_Month}
    \end{subfigure}%
    \label{fig:ALLCrimeMonthandTemp}
    \caption{Crime by month and temperature}
\end{figure}

We further explored the relation between the crime, temperature and months. From Figure \ref{fig:Temp_vs_Crime_by_Month} it is evident that during the months of June, July, and August the crime rate is highest.This could be because people go out more often during this time, are in a holiday mood and hence less careful with things, and also during this time there are more visitors int the city. However, it should be noted that from figure \ref{fig:ALLCrime_by_temperature} we found a peak of crime rate when the temperature is around $32^{\circ}$ Fahrenheit. Indeed, we found that during the months of October and November there is a rise of crime rate.   
\subsubsection{Burglary is less in regions with high property rate}
We see that there is a significant negative correlation between property price rate and burglary. A potential explanation for this is that, in a high property rate area, houses will have better safety and hence less burglary.
\begin{figure}[H]
	\centering
    \includegraphics[scale=0.3]{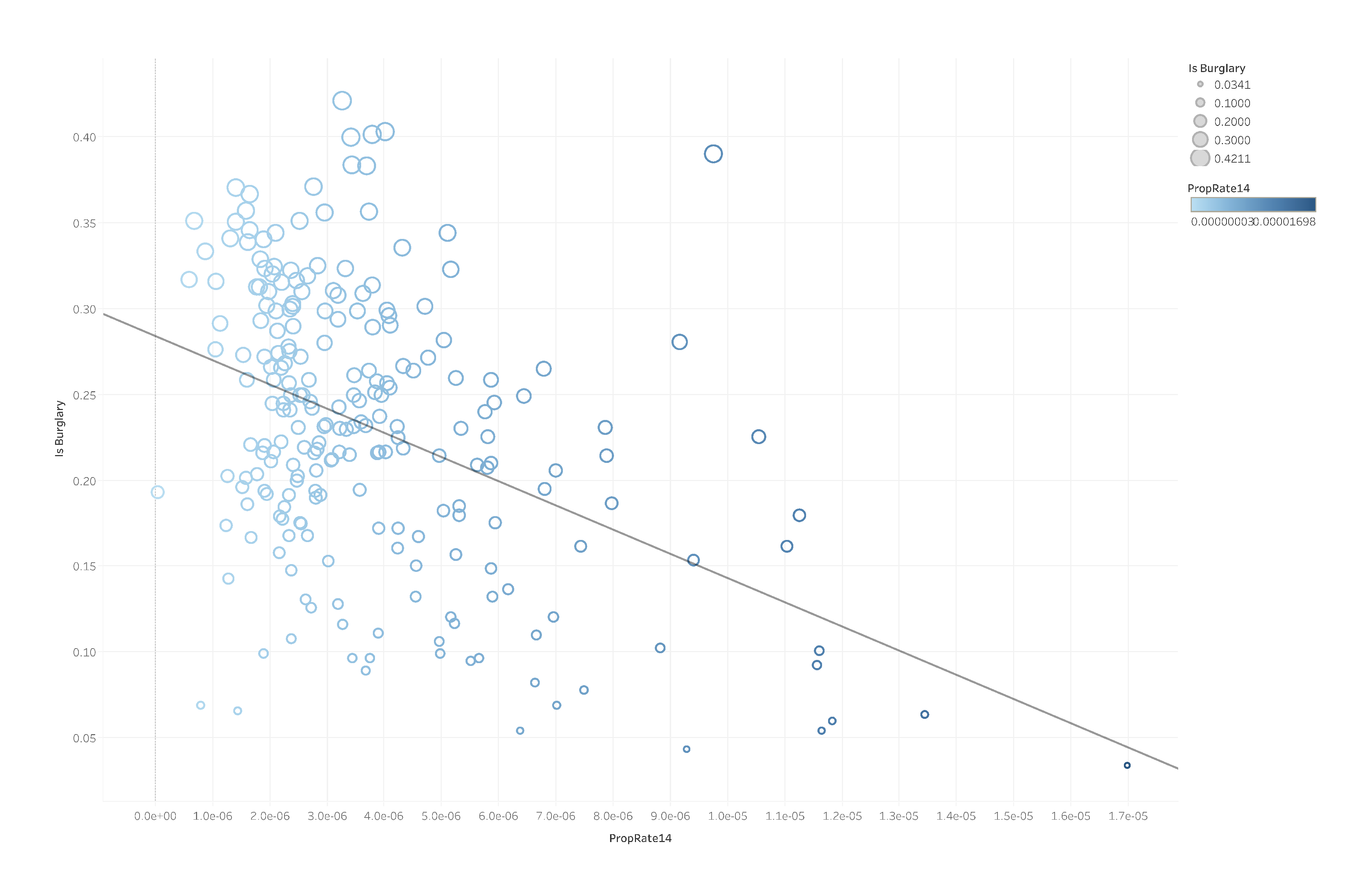}
	\caption{Burglary and Property rate Negatively correlated}
    \label{fig:propcrime}
\end{figure}
\subsection{Geographically Weighted Regression}

\subsubsection{Motivation}
Linear Regression one of the simplest methods which has a wide range of application in modeling the data. In this model the predicted variable is modeled as linear combinations of the linear features. In this model the data points are assumed to be independent of each other which might not be the case for spatial data. Hence it might not be a good method to model a spatial data. In Geographically Weighted Regression (GWR)~\cite{brunsdon1996geographically} it is assumed that the nearer data points are more correlated than the distant data points which can overcome the problem of data independence. With this motivation GWR is applied widely ~\cite{graif2009spatial,fotheringham1998geographically} to model spatial data. In the next section we will describe the model explicitly.
\subsubsection{Theory}

\underline{\textbf{The Model:}}\\

We assume that the dataset has $p$ independent variables (we will often call them as features) $X_1,\cdots, X_p$ and a dependent variable $Y$ which we want to predict. Let $(y_i;x_{i1},\cdots, x_{ip})$ is the observation of the variables at location $(u_i,v_i)$. Now the data is fit using a set of spatially varying parameters in the following form:
\begin{equation}
y_i = \sum_{j=1}^p = \beta(u_i,v_i) x_{ij} + \epsilon_i,
\end{equation}
where $\epsilon_i$ is the error parameter with mean zero and common variance $\sigma^2 . \beta(u_i,v_i)$.\\ 

\noindent \underline{\textbf{Fitting Method:}}\\

In GWR the model parameters at each location are learned using weighted least square approach. In particular if $(u_i,v_i)$ is chosen as the regression point then the weight at location $(u_j,v_j)$ ($w_j(u_i,v_i)$) is  
\begin{figure}[H]
\begin{center}
\includegraphics[scale=0.4]{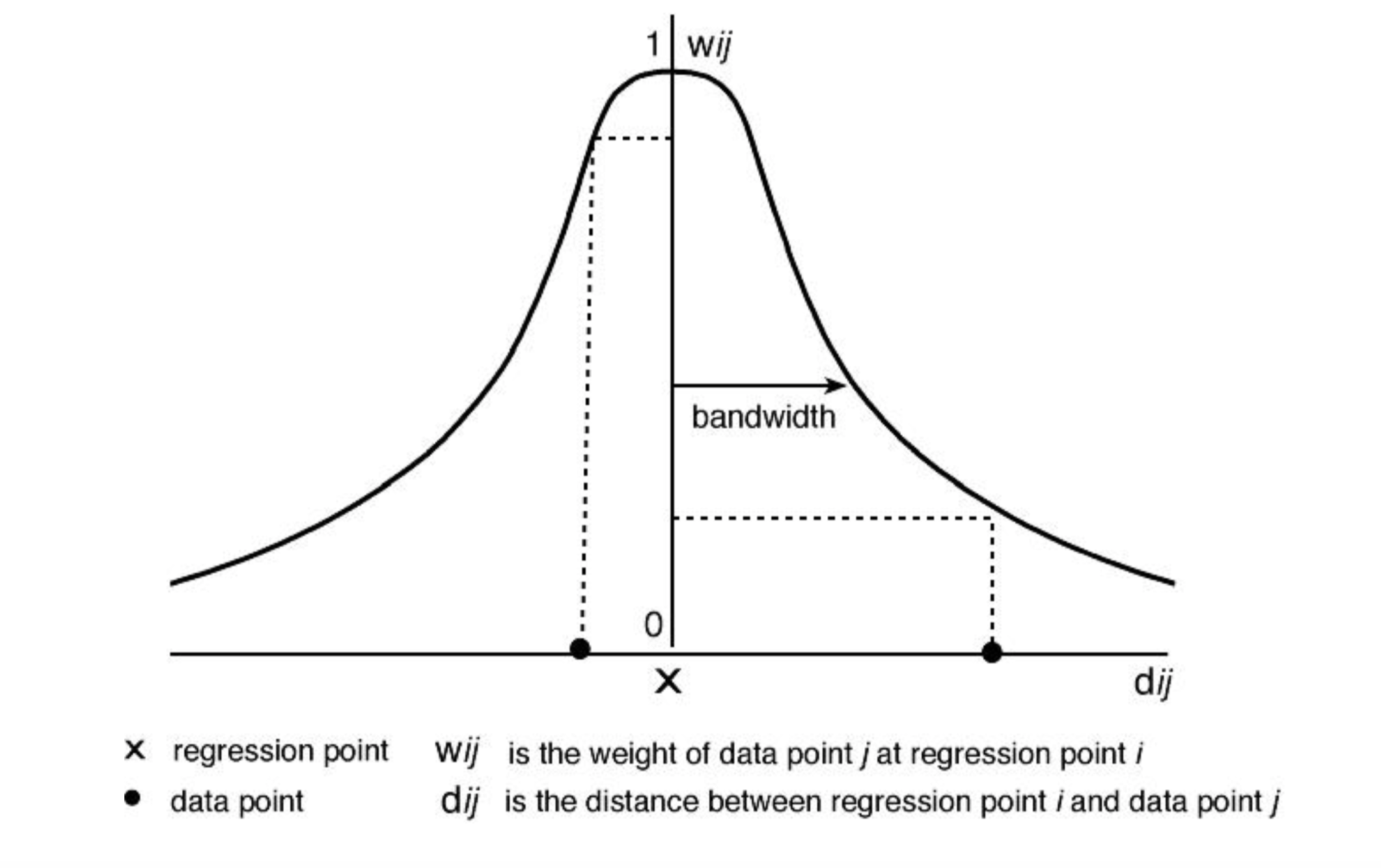}
\end{center}
\caption{Weight Matrix for GWR\cite{charlton2009geographically}}
\label{choice_weight}
\end{figure}
proportional to the distance between location $i$ and location $j$ for all $j \in [1,n]$ as shown in Figure \ref{choice_weight}. The parameters at location $(u_j,v_j)$ are then estimated by minimizing the following loss function:
\begin{equation}
\sum_{j=1}^{n} w_j(u_i,v_i) [y_j - \sum_{k=1}^p\beta(u_j,v_j) x_{jk}]
\end{equation}
It can be showed that this minimization problem has a closed form solution. Although one challenge is to choose the weight matrix. One obvious choice can be 
$$w_j(u_i,v_i) = \exp[-(d_{ij}/h)^2], j=1,\cdots,n,$$
where $d_{ij}$ is the distance between regression point $i$ and data point $j$. 
\subsubsection{GWR for Rochester Crime Dataset}
We model the crime data using the GWR Model. Our features and dependent variables are as follows:\\

\noindent \underline{\textbf{Features:}}\\

For modeling the data we have used features from all three datasets described in Section \ref{data_section}. We can divide the features as spatial and temporal features:
\begin{enumerate}
\item \textbf{Spatial Features:} We used the demographic data to extract spatial features. For example we have used the \textit{\textbf{ethnicity, age, population density, property rates} } as spatial features. Our goal is to find out how does this features of the demographics influence crime? For example using the data statistics we have seen that the places with higher property rate is less prone to crime. This is shown in Figure \ref{Proprate_vs_Burglary}.
\begin{figure}[H]
\begin{center}
\includegraphics[scale=0.4]{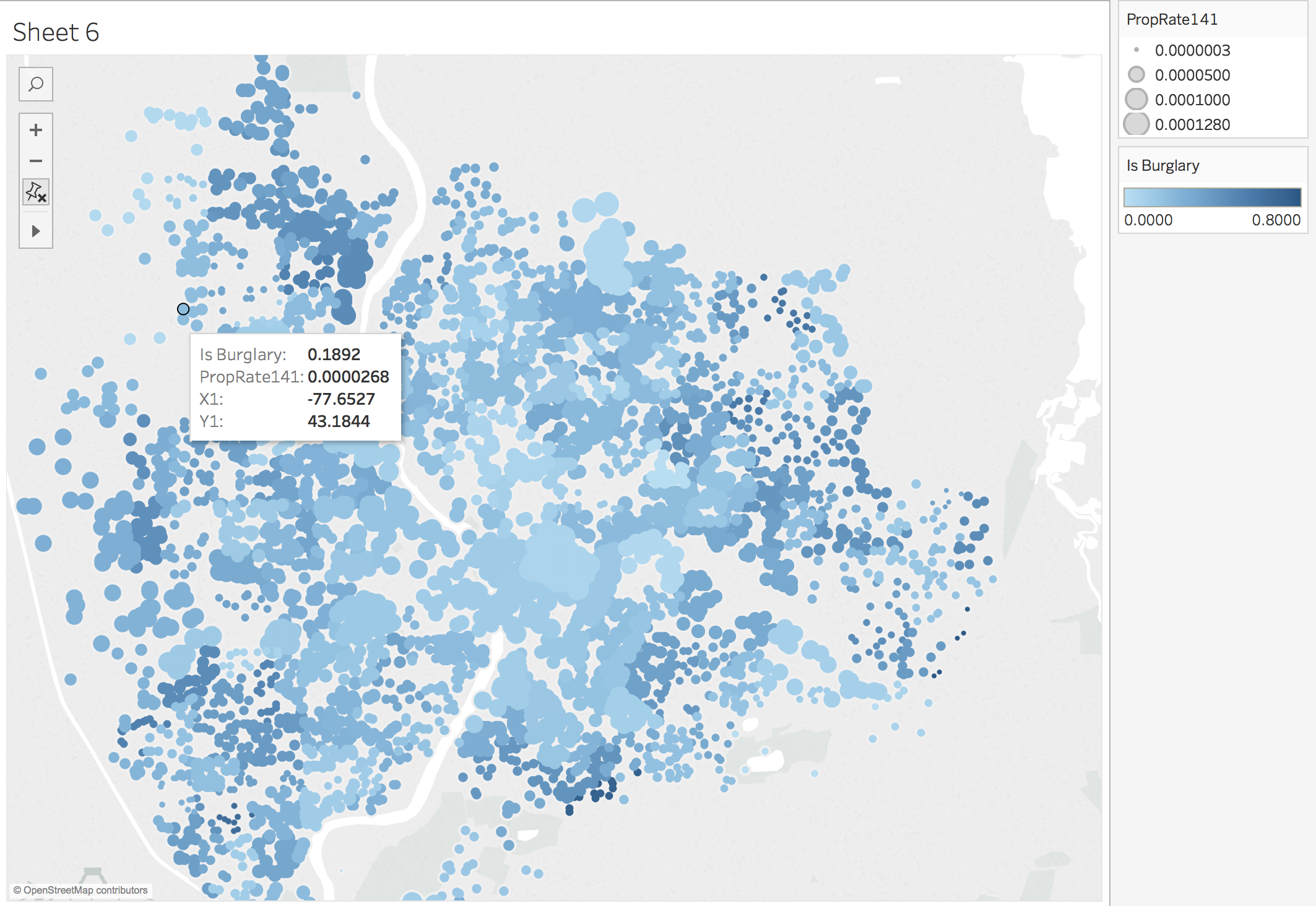}
\end{center}
\caption{Property Rate vs Probability of Burglary. \href{https://public.tableau.com/views/2017-proprate-larceny/Sheet1?:embed=y&:display_count=yes&publish=yes}{Click here for an interactive version}}
\label{Proprate_vs_Burglary}
\end{figure}

\item \textbf{Temporal Features:} We used the time of the crime to extract temporal features. We have 4 binary variables for the time depending on whether it is morning, noon or night. Also from the weather data we extracted features like the average temperatures of the day.   
\end{enumerate}

\noindent \underline{\textbf{Variables to be Predicted:}}\\

The goal of the model is to predicting crime at some given location and time. To do this we created variables for each crime type. For example we have variable \textit{\textbf{IsLarceny}} which is the probability of larceny happening at a GeoID (GeoID is unique identification number of some small geographic area and hence the probability is nonzero).

\section{Results and Discussion}

In order to test the model, we used the spatial and temporal features to predict the probability of a type of crime at different spatial locations for every year. We held out 20\% of the data points and trained the model on the remaining 80\% of the data. In order to test the model predictions, we took a new coordinate (from the test set) and obtained the corresponding feature and weight vector by averaging across the GeoID to which the point belongs. We quantify the fit by plotting the predicted against the emperical responses. Most of the data lies along the diagonal. The proportion of explained variance ($R^2$) is given next to each fit. The plots for Burglary and Larceny are given in figures 7 and 8 respectively.
\begin{figure}[H]
\begin{center}
\includegraphics[scale=0.4]{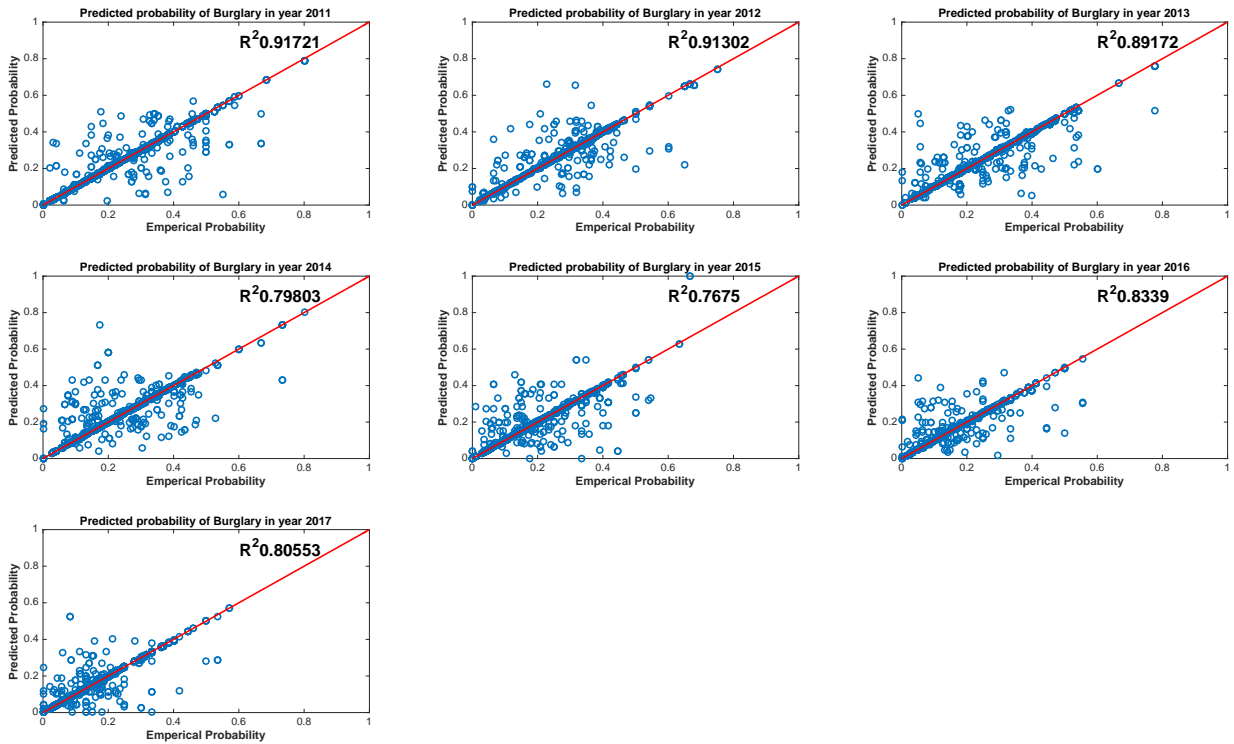}
\end{center}
\caption{Prediction of Burglary Scatter Plot for Each Year}
\end{figure}

\begin{figure}[H]
\begin{center}
\includegraphics[scale=0.4]{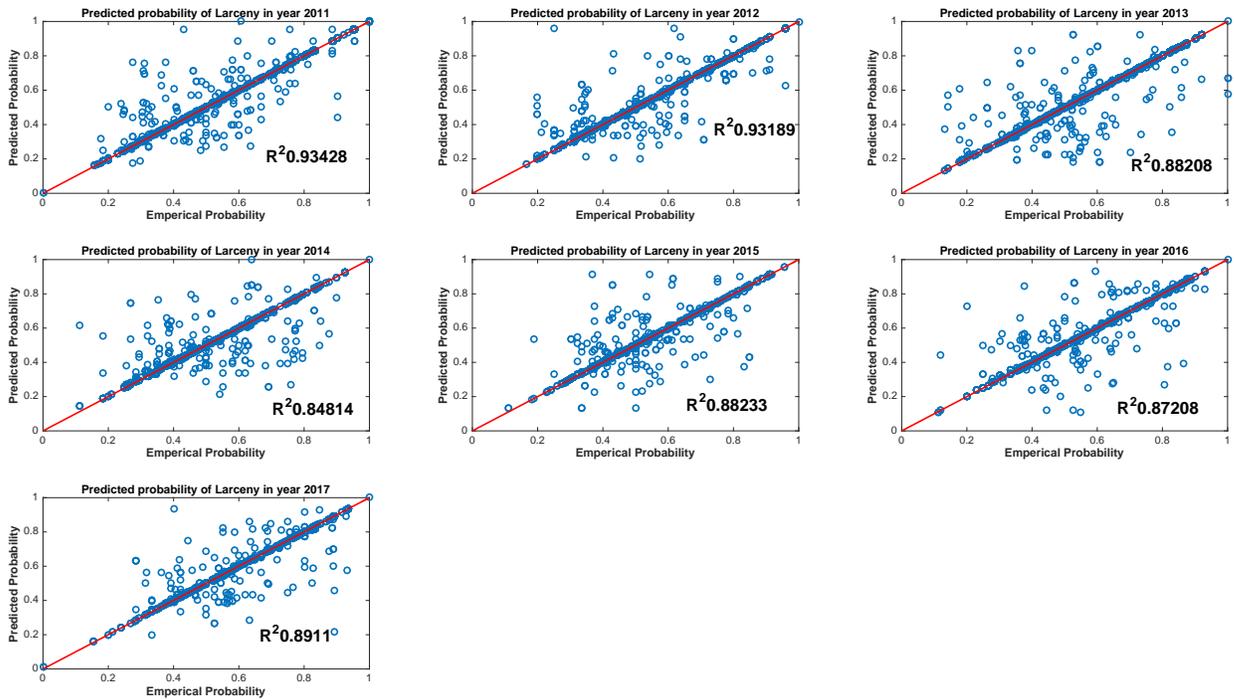}
\end{center}
\caption{Prediction of Larceny Scatter Plot for Each Year}
\end{figure}

Since the model depends on the spatial and temporal features, we can make predictions for the entire region at a particular time. We construct heat maps for the entire region and overlay it on the map of Rochester. The value at each point indicates the probability of a burglary at the particular point. An interesting observation is that the maps are relatively smooth across areas and stable over time. 

\begin{figure}[H]
\begin{center}
\includegraphics[scale=0.8]{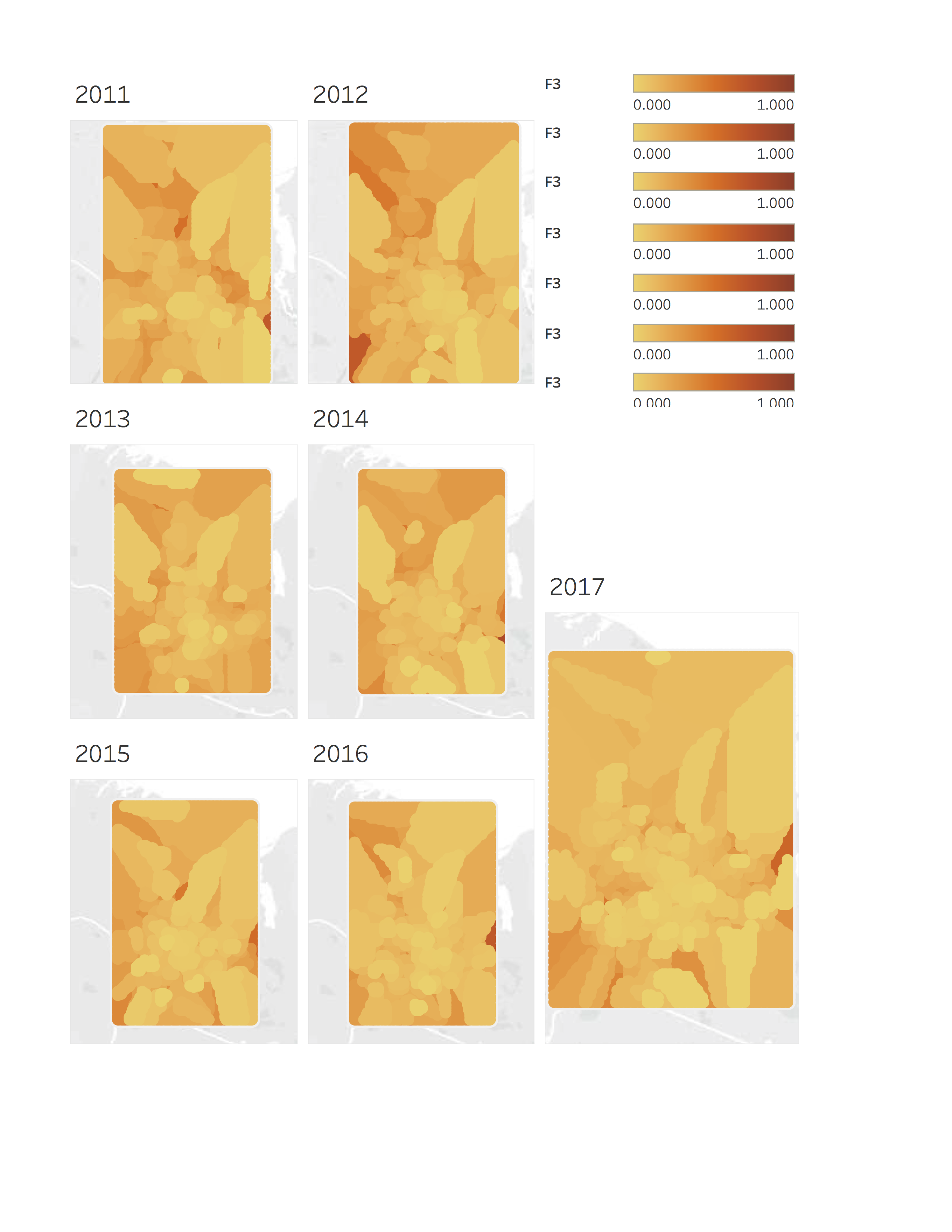}
\end{center}
\caption{Predicted Burglary Heat Map for Each Year. For an interactive version, \href{https://public.tableau.com/views/Predicted_crimerate/Dashboard1?:embed=y&:display_count=yes&publish=yes}{click here}  }
\end{figure}

\section{The Website}
To present our work and give people access to the insights we found from our analysis, we created a website (\url{https://tinyurl.com/upstatwebsite}). The website features our findings in details. The featured works in our website contains the interactive maps with crime probabilities as well as the initial data visualizations. Our website also features a \"locate me' service, which, in future, will serve as a crime occurrence indicator in a particular area. 

\section{Future Direction}
The model that we are working with is a spatial predictive model. The next step is to add a temporal component to it and make predictions dynamically. We also plan to build a generalized application that can be installed on a cellphone and used in real time to know the crime potentials in an area, based on one's GPS location. This can help a person be prepared and be more aware. Also the application can have an option such that it can be clicked and the local police can be informed when the person feels insecure.
\newpage
\bibliography{crime}
\bibliographystyle{plain}

\newpage
\appendix
\section{APPENDIX}
\subsection{Crime by time of day}
\begin{figure}[H]
    \centering
    \begin{subfigure}[b]{0.5\textwidth}
        \includegraphics[scale=0.4]{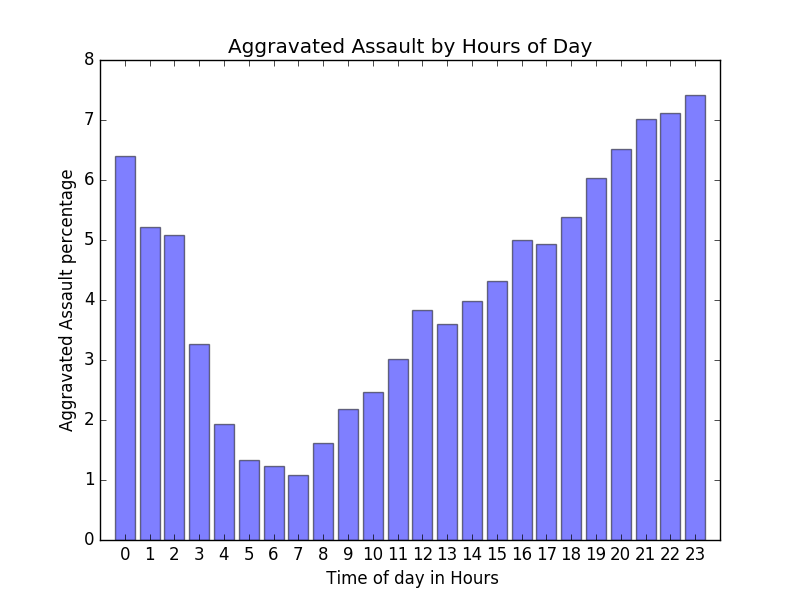}
        \caption{Aggravated Assault}
        \label{fig:Aggravated_AssaultbyHoursofDay}
    \end{subfigure}%
    \begin{subfigure}[b]{0.5\textwidth}
        \includegraphics[scale=0.4]{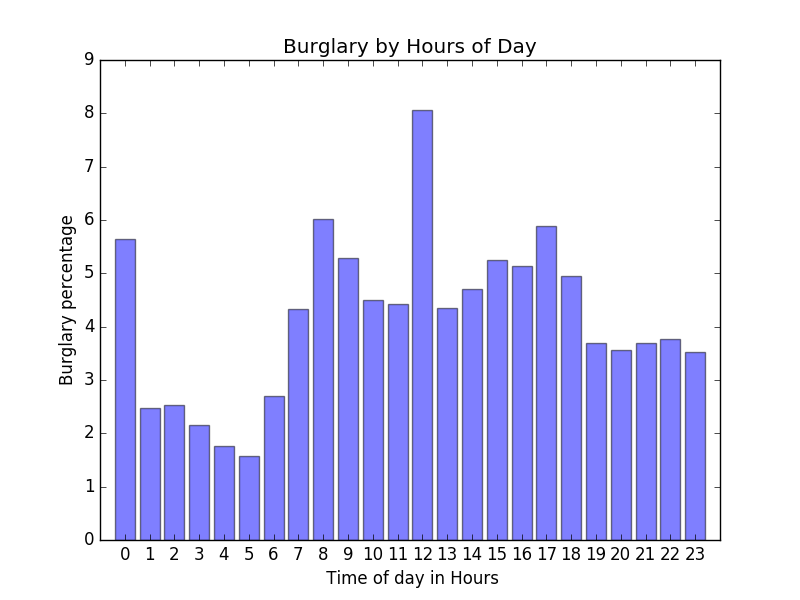}
        \caption{Burglary}
        \label{fig:BurglarybyHoursofDay}
    \end{subfigure}%
    \\
    \begin{subfigure}[b]{0.5\textwidth}
        \includegraphics[scale=0.4]{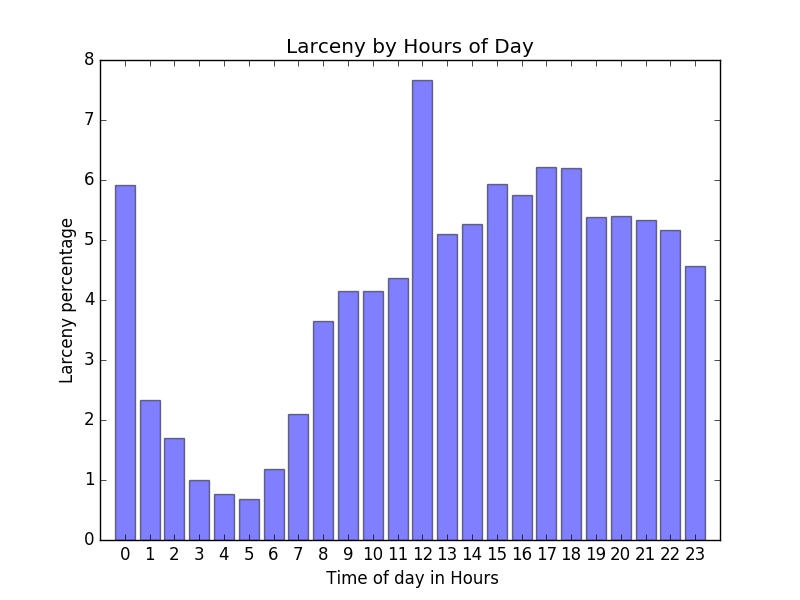}
        \caption{Larceny}
        \label{fig:LarcenybyHoursofDay}
    \end{subfigure}%
    \begin{subfigure}[b]{0.5\textwidth}
        \includegraphics[scale=0.4]{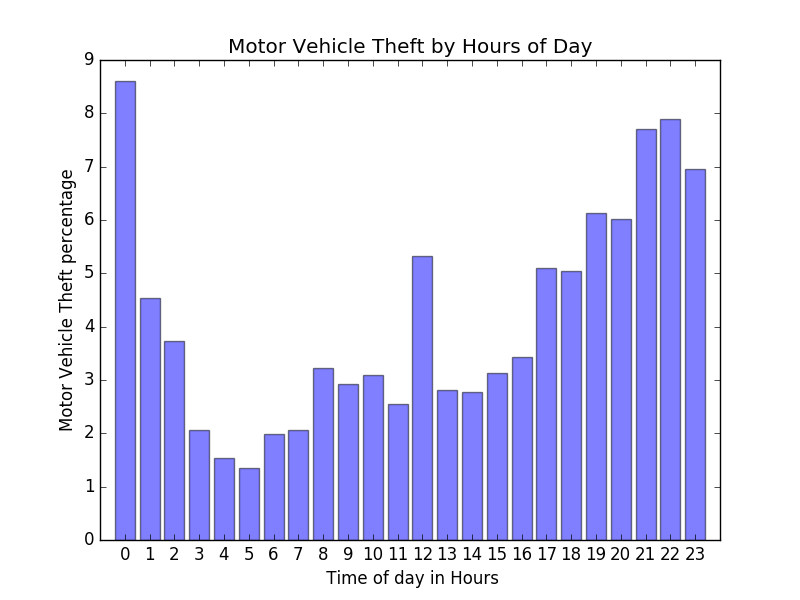}
        \caption{Motor Vehicle Theft}
        \label{fig:Motor_Vehicle_TheftbyHoursofDay}
    \end{subfigure}%
    \\
    \begin{subfigure}[b]{0.5\textwidth}
        \includegraphics[scale=0.4]{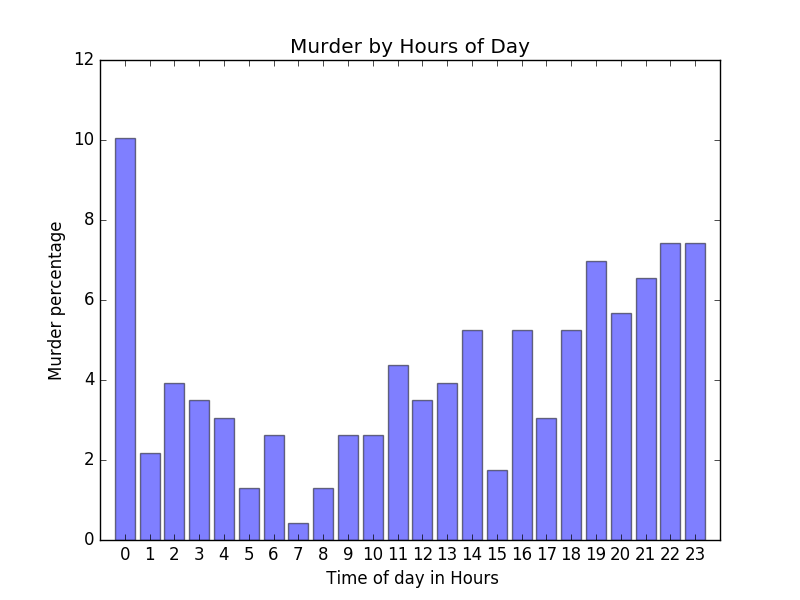}
        \caption{Murder}
        \label{fig:MurderbyHoursofDay}
    \end{subfigure}%
    \begin{subfigure}[b]{0.5\textwidth}
        \includegraphics[scale=0.4]{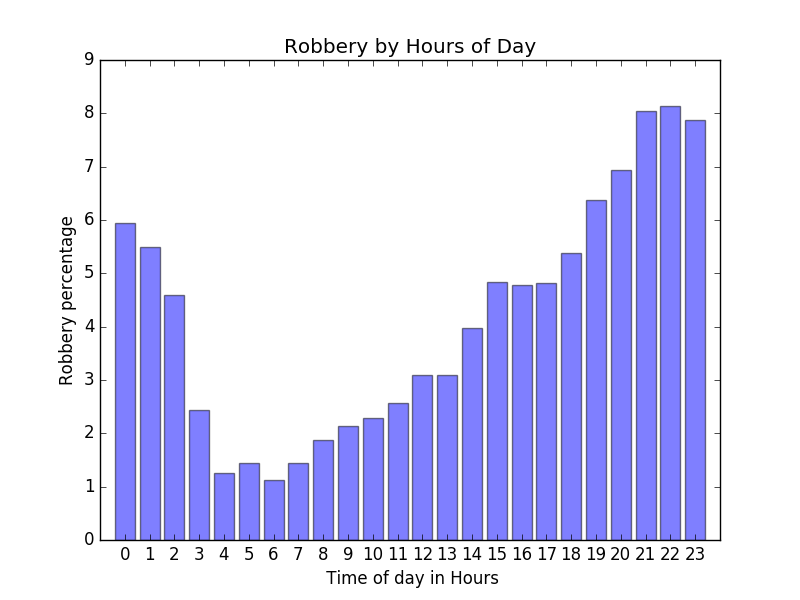}
        \caption{Robbery}
        \label{fig:RobberybyHoursofDay}
    \end{subfigure}%
    \label{fig:Crimebyday}
	\caption{Crime by time of day}
\end{figure}

\subsection{Crime by Temperature}
\begin{figure}[H]
    \centering
     \begin{subfigure}[b]{0.5\textwidth}
        \includegraphics[scale=0.4]{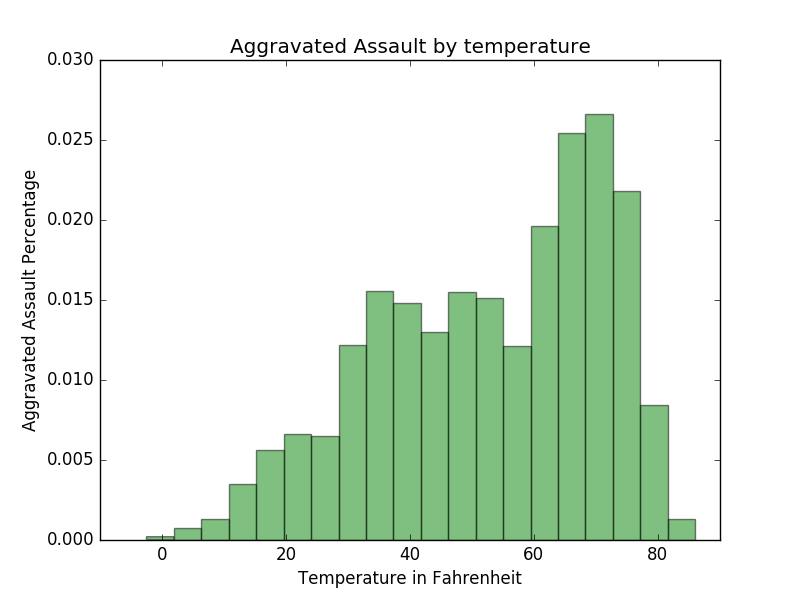}
        \caption{Aggravated Assault}
        \label{fig:Aggravated_Assault_by_temperature}
    \end{subfigure}%
    \begin{subfigure}[b]{0.5\textwidth}
        \includegraphics[scale=0.4]{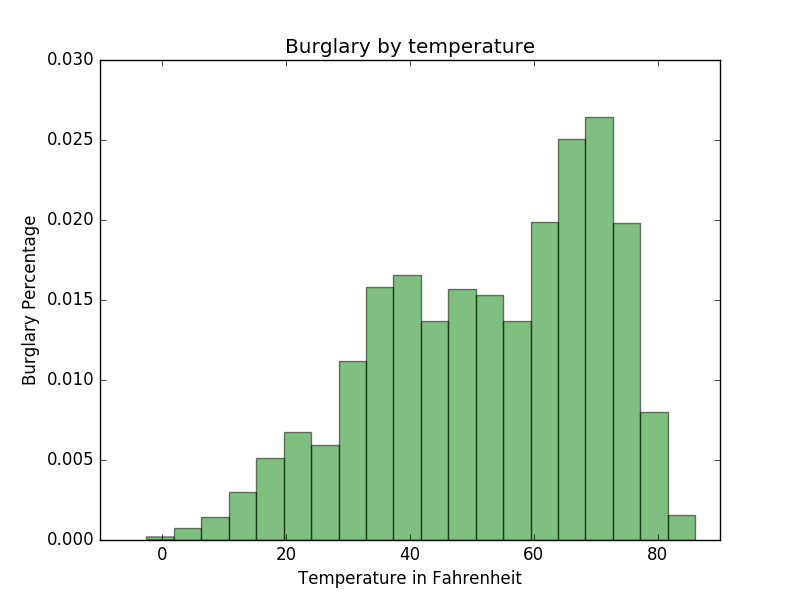}
        \caption{Burglary}
        \label{fig:Burglary_by_temperature}
    \end{subfigure}%
    \\
    \begin{subfigure}[b]{0.5\textwidth}
        \includegraphics[scale=0.4]{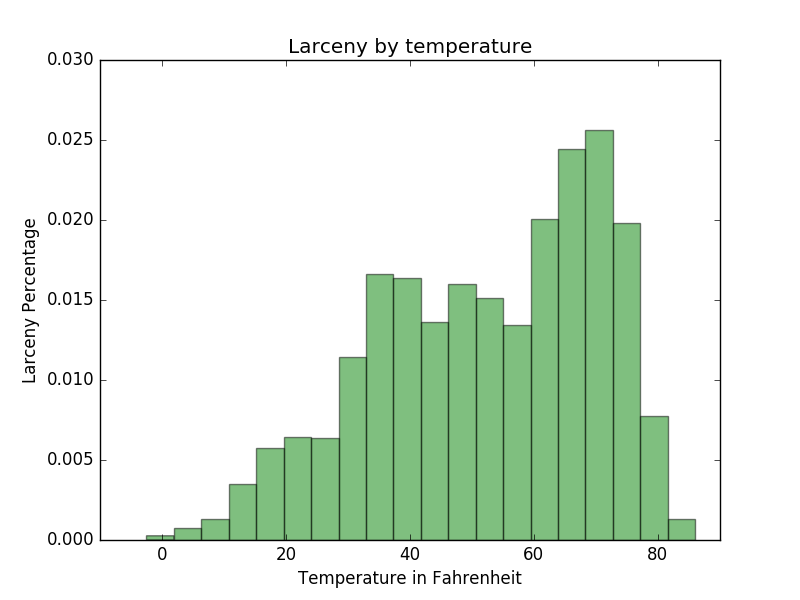}
        \caption{Larceny}
        \label{fig:Larceny_by_temperature}
    \end{subfigure}%
    \begin{subfigure}[b]{0.5\textwidth}
        \includegraphics[scale=0.4]{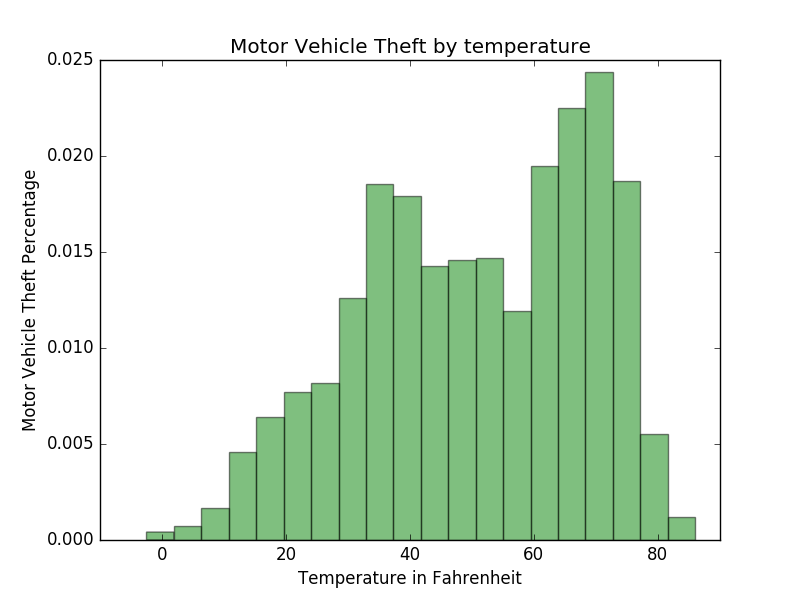}
        \caption{Motor Vehicle Theft}
        \label{fig:Motor_Vehicle_Theft_by_temperature}
    \end{subfigure}%
    \\
    \begin{subfigure}[b]{0.5\textwidth}
        \includegraphics[scale=0.4]{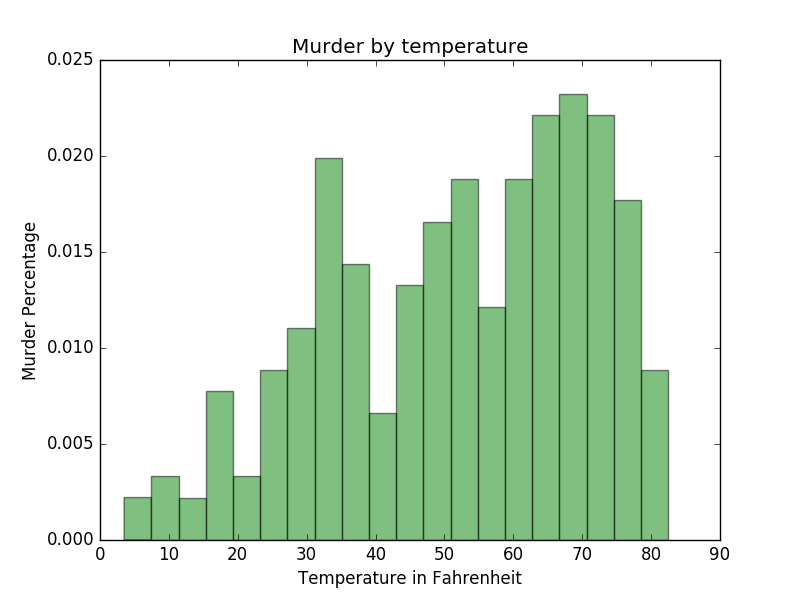}
        \caption{Murder}
        \label{fig:Murder_by_temperature}
    \end{subfigure}%
    \begin{subfigure}[b]{0.5\textwidth}
        \includegraphics[scale=0.4]{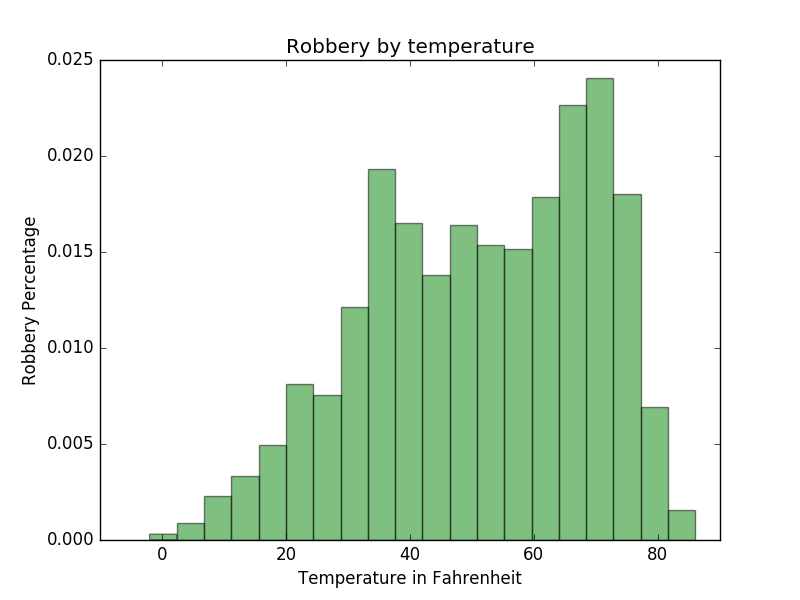}
        \caption{Robbery}
        \label{fig:Robbery_by_temperature}
    \end{subfigure}%
	\caption{Crime by temperature}
    \label{fig:Crime_by_temperature}
\end{figure}

\subsection{Crime by Month}
\begin{figure}[H]
    \centering
     \begin{subfigure}[b]{0.5\textwidth}
        \includegraphics[scale=0.4]{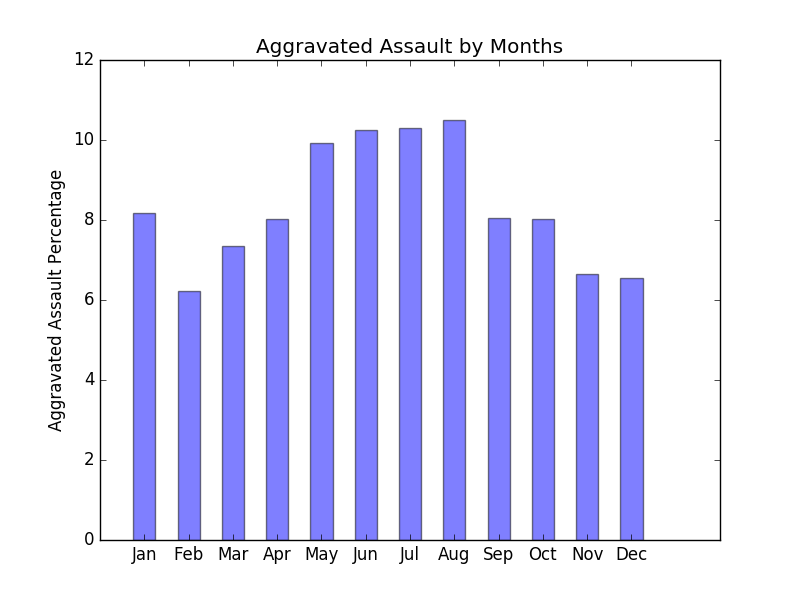}
        \caption{Aggravated Assault}
        \label{fig:Aggravated_Assault_by_Months}
    \end{subfigure}%
    \begin{subfigure}[b]{0.5\textwidth}
        \includegraphics[scale=0.4]{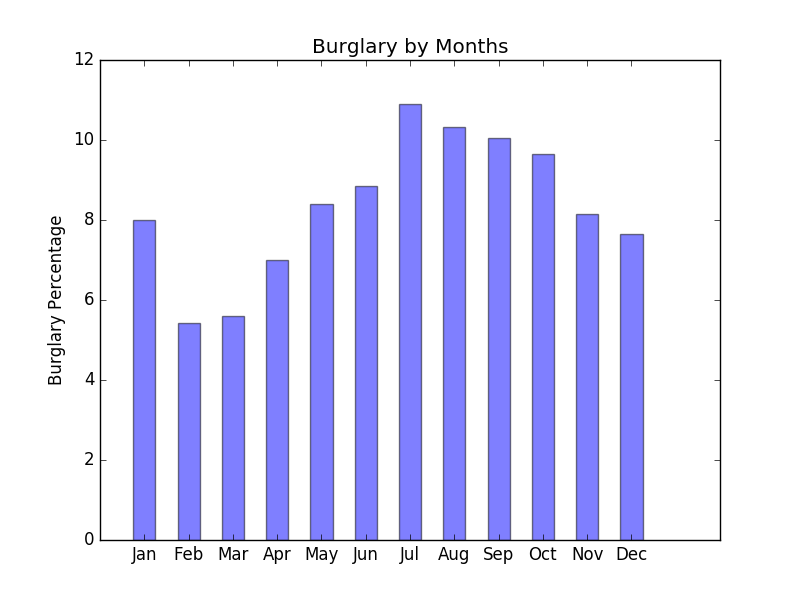}
        \caption{Burglary}
        \label{fig:Burglary_by_Months}
    \end{subfigure}%
    \\
    \begin{subfigure}[b]{0.5\textwidth}
        \includegraphics[scale=0.4]{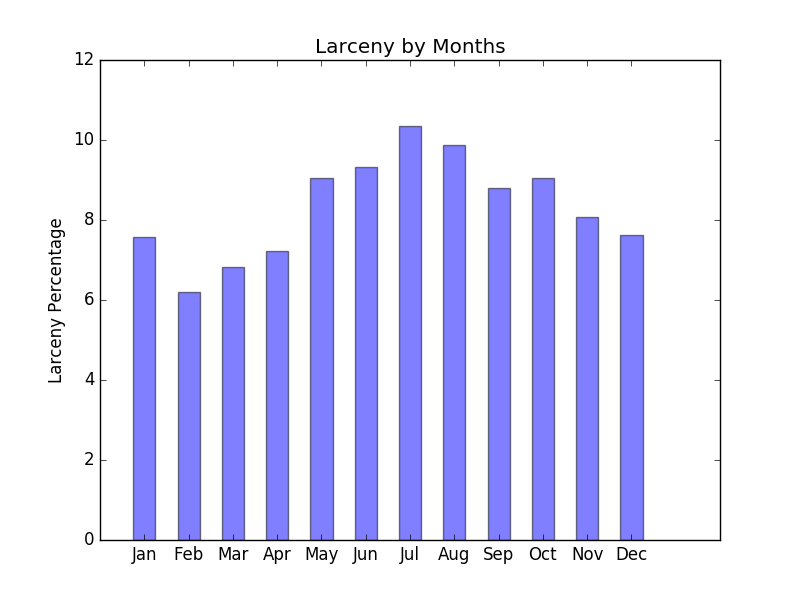}
        \caption{Larceny}
        \label{fig:Larceny_by_Months}
    \end{subfigure}%
    \begin{subfigure}[b]{0.5\textwidth}
        \includegraphics[scale=0.4]{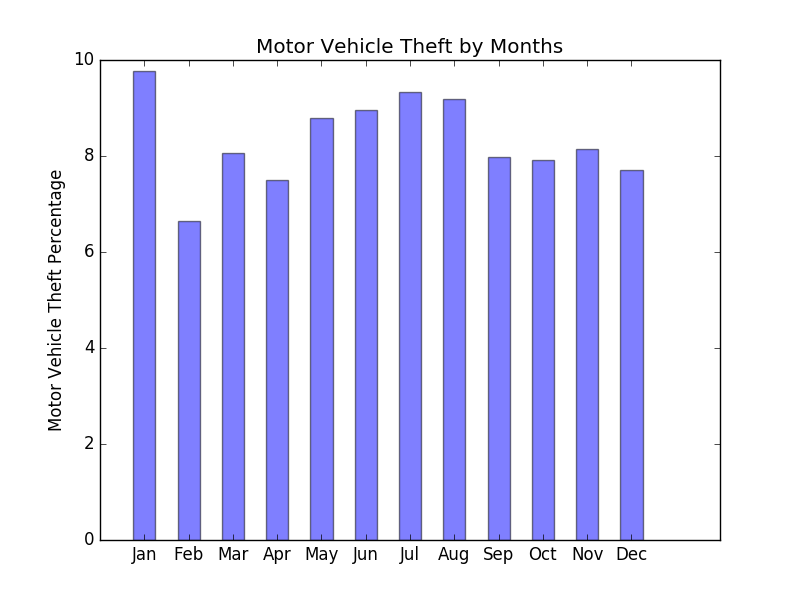}
        \caption{Motor Vehicle Theft}
        \label{fig:Motor_Vehicle_Theft_by_Months}
    \end{subfigure}%
    \\
    \begin{subfigure}[b]{0.5\textwidth}
        \includegraphics[scale=0.4]{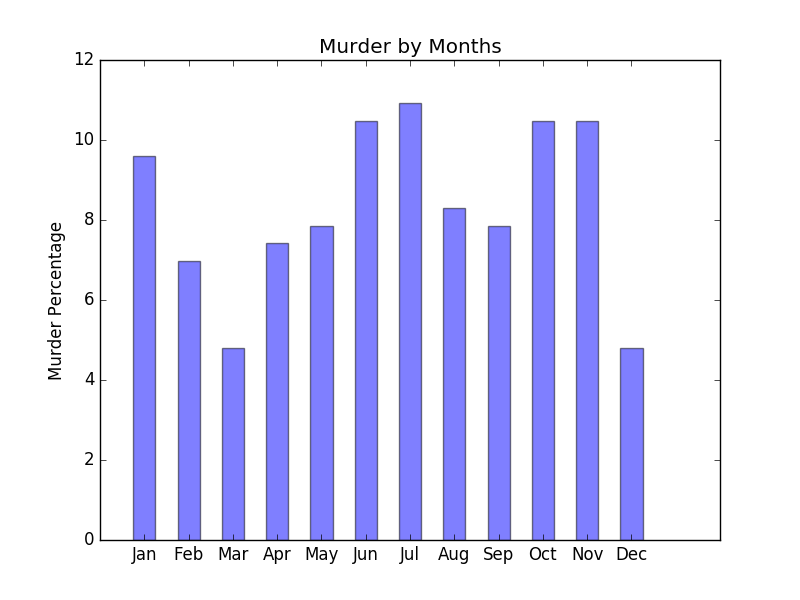}
        \caption{Murder}
        \label{fig:Murder_by_Months}
    \end{subfigure}%
    \begin{subfigure}[b]{0.5\textwidth}
        \includegraphics[scale=0.4]{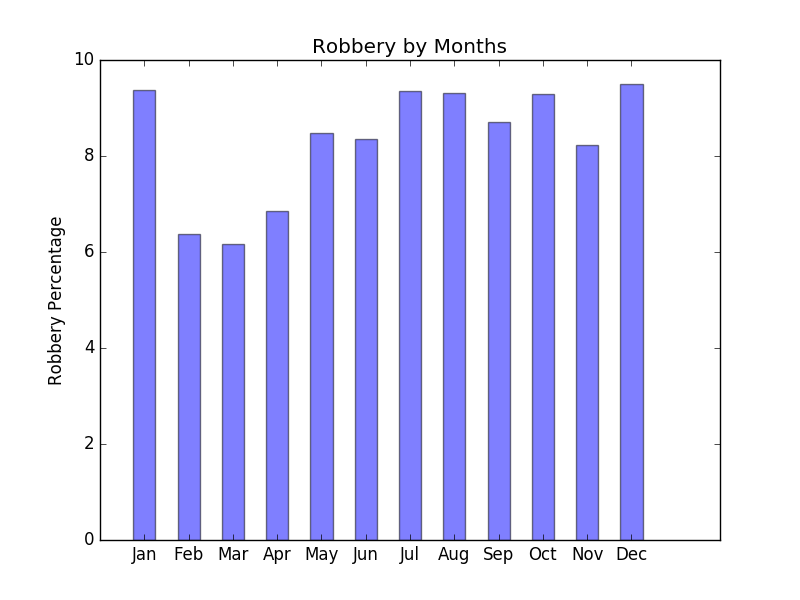}
        \caption{Robbery}
        \label{fig:Robbery_by_Months}
    \end{subfigure}%
	\caption{Crime by temperature}
    \label{fig:Crimes_by_Months}
\end{figure}
\end{document}